\documentclass[a4paper]{article}
\usepackage[utf8]{inputenc}
\usepackage[russian]{babel}
\usepackage{amsmath}
\usepackage{amssymb,amsfonts,textcomp}
\usepackage{color}
\usepackage{array}
\usepackage{hhline}
\usepackage{hyperref}
\usepackage[pdftex]{graphicx}
\usepackage{listings}

\title{Оптимальные параметры для соединения данных на Apache Spark с фильтрами Блума}
\author{Ophir LOJKINE}
\date{2017-06-09}
\begin{document}
\title{Оптимальные параметры для соединения данных на Apache Spark с фильтрами Блума}
\maketitle

\section{Аннотация}
В этой статье мы представим такой алгоритм соединения таблиц в кластере, который позволит быстро фильтровать данные до
соединения, чтобы ускорить соединение в случае, если ода из таблиц больше другой, а результат соединения включает не
много записей. 

\section{Введение}
Схемы баз данных типа «Звезда» или «Снежинка» используют одну большую таблицу фактов и несколько маленьких таблиц
измерения. Подобные схемы часто требуют фильтровки в таблицах измерения, поэтому такие схемы требуют  обработки
множества записей даже когда результат запроса маленький по объему. Наша работа не затрагивает исключительно подобные
таблицы. В этой статье мы предположим, что у нас имеется только две таблицы, одна из которых по объему больше другой.
Одна из таблиц достаточно маленькая (в статье будет раскрыто понятие «достаточно»). Другая таблица условно будет
называться большой. Обе таблицы распределенные и находятся на одном кластере. Цель данного научного исследования —
выполнить следующий запрос: 

\begin{lstlisting}[language=SQL]
SELECT
  BIGTABLE.attribute1, SMALLTABLE.attribute2
FROM
  BIGTABLE INNER JOIN
  SMALLTABLE ON BIGTABLE.key = SMALLTABLE.key
WHERE
      condition1(BIGTABLE.attribute3)
  AND condition2(SMALLTABLE.attribute4)
\end{lstlisting}

\section{Степень научной изученности}
Наша работа в большинстве своем основана на исследованиях \textit{Brito et al., 2007}. Эти исследователи предложили два
алгоритма для быстрого соединения данных в случае, который нас интересует. Один из них теперь включен по умолчанию в
Spark и о нем уже много писали. Этот алгоритм называли «SBJ» (Spark Broadcast join). В Spark она называется  Broadcast
hash join. 

Но другой алгоритм является новым и достаточно мало изучен. Его назвали «SBFCJ» (Spark Bloom-filtered cascade join).
Суть алгоритма довольно проста - это создание фильтра Блума, который будет содержать ключи маленькой таблицы и
использовать этот фильтр, чтобы уменьшить количество записей в большой таблицы до соединения. Все ключи маленькой
таблицы отправляются в управляющий узел кластера, затем управляющий узел создает за один раз фильтр Блума определенного
размера, отправляет этот фильтр всем остальным узлам с помощью механизма Broadcast. Все узлы фильтруют большую таблицу,
убирая элементы которых нет в фильтре. В конце, отфильтрованные таблицы Spark соединяет обычным образом. С момента
этого исследования в Spark появились новые механизмы, которые помогут нам предложить вариант этого подхода, который
работает быстрее.

Авторы статьи \textit{Brito et al., 2007} работали в первой версией Spark и использовали структуру данных RDD. В
полученном результате они обнаружили, что выполнение запросов для маленьких объемов данных занимает почти одинаковое
время. Это значит, что какое-либо вычисление, время которого не сильно  зависит от объема данных, занимает большую
часть времени. 

\section[Новые технологии в Spark 2]{Новые технологии в Spark 2}
\subsection{Распределённые Фильтры Блума}
Spark 2 по умолчанию предлагает слитные фильтры Блума. То есть, структура данных создаётся одновременно для всех
разделов данных живой памяти всех узлов, затем постепенно платформа соединяет фильтры. В конце концов она отправляет
конечный фильтр управляющему узлу. Это не только работает быстрее, чем создание фильтра на управляющем узле за один
раз, но и не требует того, чтобы все элементы фильтра входили в память одновременно.

\subsection{Генерация кода}
Spark 2 также предлагает оптимизатор запросов по имени Catalyst, который умеет создавать машинный код для виртуальной
машины JVM во время выполнения запроса и структуры данных (DataSet), для которых сериализация и десериализация работает
быстрее и которые он умеет сортировать без десериализации. 

\subsection[Влияние на результаты экспериментов]{Влияние на результаты экспериментов}
С одной стороны новые технологии помогут нам предложить новый более мощный вариант, вышеописанного алгоритма. Но с
другой стороны, механизм выполнения запросов по умолчанию работает иначе и быстрее и, возможно, нам не понадобится
другой алгоритм, так как тот, что есть по умолчанию, работает  быстрее. Это то, что мы уточним в данной статье с
помощью Spark 2, проведя ряд экспериментов на разных кластерах. В последней части статьи мы предложим упрощенную
математическую модель времени выполнения интересующих нас запросов, которая примерно предупреждает время выполнения
запроса в зависимости от топологии и параметров кластера и от свойств таблиц.

\section[Предлагаемый алгоритм ]{Предлагаемый алгоритм }
\subsection{Предлагаемые изменения традиционного соединения с фильтром Блума}
Мы предлагаем два основных изменения этого алгоритма. 

\begin{itemize}
\item Вычислить фильтр Блума не на управляющем узле, а на всех узлах параллельно, потом слить их, используя свойства
фильтра Блума, который позволяет совершать быстрое вычисление и слияние. 
\item Создать фильтры не определенного размера, а такие, размер которых зависит от количества записей в маленькой
таблице. Для этого мы используем метод Spark,  который позволяет  получить примерное значение количества элементов в
таблице данных, затрачивая определенное время. 
\end{itemize}
К тому же, мы переделали программу бразильских исследователей на основе новых API в Spark 2. Следовательно, наша
программа затрачивает намного меньше времени на сериализацию и десериализацию данных, так как API DataSet в Spark 2 не
требует десериализации элементов во время этапа Shuffle. Также автоматическая генерация кода в новой версии уменьшает
количество этапов выполнения программы. 

\subsection{Алгоритм }
Первый шаг: быстрое вычисление размера маленькой таблицы. Используя механизм Spark, который умеет возвращать не конечный
 темпоральный результат до конца одной задачи, мы тратим определенное количество секунд на получение примерного
значения размера маленькой таблицы. 

Второй шаг: определение параметров фильтра Блума. 

Чтобы создать оптимальный фильтр Блума, надо задавать два из этих трех параметров: 

\begin{itemize}
\item количество элементов, которые мы будем добавлять в фильтр 
\item размер фильтра в битах
\item желательный процент ошибок (вероятность того, что фильтр выдаст ложно-положительный результат во время запроса)
\end{itemize}
Мы задаем два параметра. Это количество элементов, число которых мы знаем, благодаря первому шагу и желательный процент
ошибок. В последней части мы объясним, как рассчитать желательный процент ошибок. 

Третий шаг: отправка фильтра всем узлам кластера. Как и в алгоритме оригинальной статьи, мы используем мощный механизм
Broadcast Spark, который обменивается данными с помощью протокола peer-to-peer.

Четвертый шаг: фильтровка большой таблицы с помощью отправленного фильтра. Используя метод Dataset.filter() мы уберем
записи, которых нет в маленькой таблицы, из большой таблицы. Так как фильтр Блума может выдавать ложно-положительные
результаты, несколько записей, которые отсутствуют в маленькой таблице, могут остаться в большой. Их количество зависит
от параметров, которые мы описывали во втором шаге. 

Последний, пятый шаг: соединение таблиц обычным механизмом Dataset.join(). Мы позволяем Spark  решить какой алгоритм
использовать для соединения. Если данные маленькой таблицы достаточно объемные, Spark будет использовать алгоритм Sort
merge join. То есть он будет сортировать данные со стороны этапа Map и затем они будут обмениваться в кластере и
записи, которые соответствуют в большой и маленькой таблицах будут объединены со стороны этапа Reduce.

\section{Живые эксперименты}
Мы реализовали алгоритм, описанный выше, на языке Scala. Мы его протестировали на множестве соединенных компьютерах,
которые принадлежат проекту Grid5000. Мы получили доступ к множеству разных компьютеров, и попробовали произвести наши
эксперименты на компьютерах, которые соответствуют условиям реальной жизни. 

\subsection{Используемые данные}
Мы провели наши тестирования, используя стандартные данные. Мы сгенерировали наши данные с программой TPCH-DBGEN.
Которая производит данные, согласно стандарту баз данных TPC-H. В наших экспериментах мы использовали три разных
размера данных (Scale factor), чтобы протестировать влияние размера данных соединения на алгоритм: 10, 100, 150. Мы
протестировали только соединение двух таблиц: таблица заказов (Orders) с таблицей элементов заказов (Lineitem).

Данные были сгенерированы с помощью программы Dbgen в формате CSV. Затем мы их сконвертировали из формата CSV в формат
Apache Parquet со всеми параметрами по умолчанию Spark. Это значит, что файлы  Parquet разбиты на части и каждая часть
соответствует 128 Мб в CSV. Затем мы загрузили эти файлы на распределенную файловую систему HDFS, которая есть на всех
узлах нашего кластера. 

\subsection{Программа}
Мы реализовали вышеописанный алгоритм на языке Scala. Мы также реализовали в этой же программе соединение с помощью
Spark SQL, чтобы сравнить наш подход с методом соединения Spark по умолчанию. Нашу программу мы опубликовали в
свободном доступе в Github. 

Программа работает на Spark 2.1.0 и использует программу, управляющую ресурсами YARN и HDFS из Hadoop 2.8.0. Все
параметры мы оставили по умолчанию, кроме: 

\begin{itemize}
\item количество исполнителей
\item параллелизм каждого исполнителя
\item объем живой памяти, доступной каждому исполнителю и управляющей программе
\item максимальный объем результатов задач : для самых точных фильтров Блума фильтр может быть очень объемным и по
умолчанию  Spark останавливает программу при получении таких объемных данных от задач. Мы определили этот параметр на
0, чтобы  Spark не проверил объем полученных данных. Но так как мы увидим далее, в результате нашего моделирования, мы
уточнили оптимальный размер фильтра Блума, так как для подобного размера можно использовать параметры Spark по
умолчанию. 
\end{itemize}
Важно отметить, что мы не изменили количество разделов после операции соединения по умолчанию:  Spark использует 200
разделов, и следовательно, 200 задач на последнем этапе, чтобы рассчитать результат соединения. 

Наша программа принимает в качестве параметра программы процент ложно-положительных ошибок фильтра Блума. Мы провели
несколько экспериментов с разными значениями этого параметра. 

\subsection{Результаты экспериментов}
\subsubsection{Недостатки нашего подхода}
Так как доступный нам объем долгосрочного хранилища на кластера ограничен мы не смогли сгенерировать данные с большим
размером, чем SF = 150. Этот размер слишком мал для такого мощного кластера, какой мы использовали.  Время выполнения
программы всегда было меньше 20 минут. Это значит, что время, которое Spark уделяет на инициализацию и между задачами
достаточно значительно, по сравнению с временем, которое уделено на само выполнение задач. 

Мы провели эксперименты только с управляющим ресурсами  YARN, но не с управляющим ресурсами по умолчанию  Spark.
Управляющий ресурсами может играть важную роль в распределении задач и, следовательно, это влияет на время выполнения
всей программы. 

\subsubsection{Результаты}
Ниже приведён график результатов  69 экспериментов с разными значениями процента ложно-положительных ошибок. На графике,
для каждого эксперимента указаны две точки: время выполнения этапа распределенного формирования фильтра Блума и время
фильтровки большой таблицы и самого соединения таблиц.

\begin{figure}
  \includegraphics[width=\linewidth]{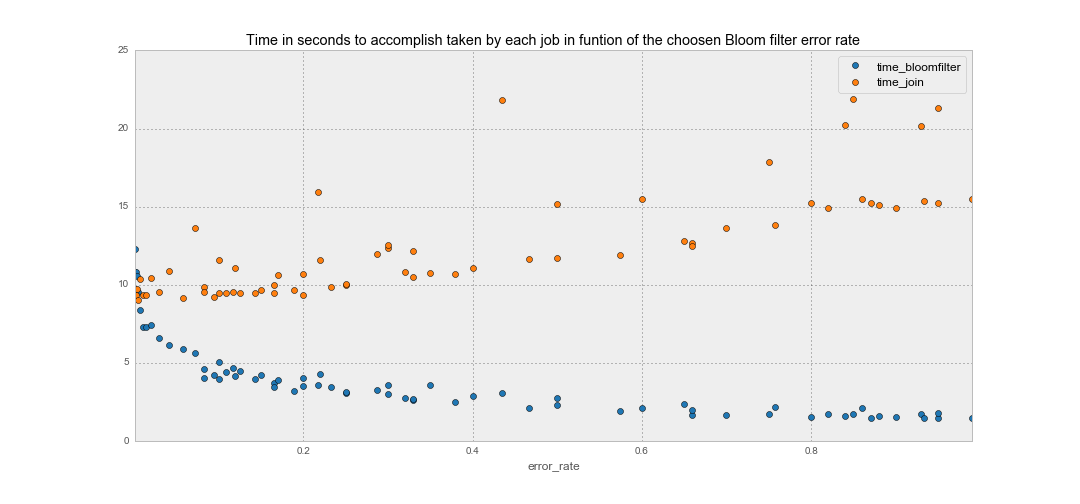}
\end{figure}
\subsubsection[Анализ результатов]{Анализ результатов}
\begin{figure}
\includegraphics[width=\linewidth]{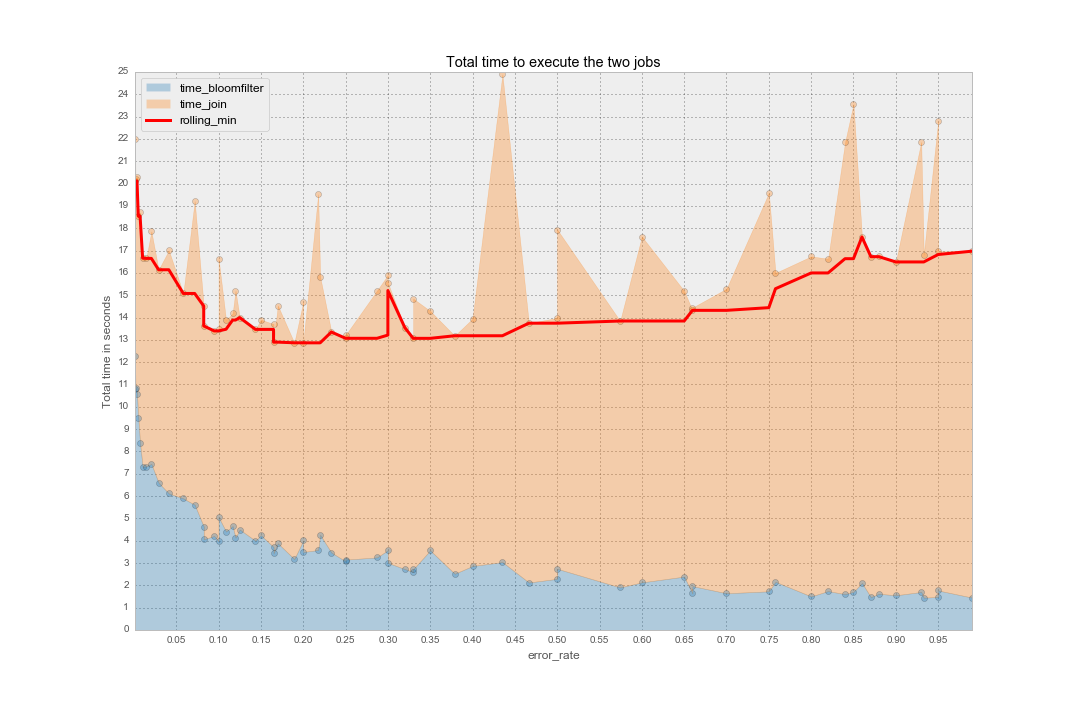}
\end{figure}
Первое, что мы видим, это то, что последний этап занимает, в большинстве случаев, намного больше времени, чем второй.
Это значительный плюс для нас.  Это для нас значит, что этап, который мы добавили не добавляет слишком много времени.
Можно предположить, что если большая таблица была бы объемнее, разница между временем этих двух этапов была бы больше:
формирование фильтра Блума зависит только от маленькой таблицы.

Второе, что мы видим, это то, что время выполнения второго этапа не регулярно. Мы это от части объяснили в детальном
анализе на английском.

Мы можем сказать по поводу наблюдаемых тенденций то, что время выполнения первого этапа очень большое для фильтров Блума
с процентом ошибок менее 5\%. В части этой статью, где рассмотрена модель, мы это объясняем. 

Мы также будем предлагать, примерную математическую модель времени выполнения второго этапа, в зависимости от количества
отфильтрованных элементов фильтром Блума, и от процента ошибок  фильтра.

\section[Модель]{Модель}
Модель мы создали в нашем детальным анализом результатов на английском.

Детальный анализ на ангиском можно найти на следующим адресом: 
\url{https://github.com/lovasoa/spark-bloomfiltered-join-analysis/blob/master/analysis.ipynb}

\subsection{Итоги анализа}
\subsubsection{Время формирования фильтра}
Время выполнения первого этапа имеет следующую форму:

\begin{equation*}
\mathit{bloomCreationTime}=K_1\times \mathit{bloomFilterSize}+K_2
\end{equation*}
где

\begin{itemize}
\item $K_2$ – постоянный элемент: время которое платформа тратит не на обработке битов наших фильтров
Блума.
\item $K_1$ – постоянное время, которое платформа тратит на каждый бит фильтра Блума. Платформа Spark
должна обменивать фильтры Блума между узлами, и это занимает время, которое линеарно зависит от размера фильтра в
байтах. Платформа тоже должна слить все фильтры Блума которые она создала. Это операция довольно простая: это бинарная
операция логического сложения (дизъюнкция) на биты наших фильтров Блума. Время выполнения логического сложения на двух
массивах одинакового размера занимает время, которое зависит линеарно от размера.
\item \begin{equation*}
\mathit{bloomFilterSize}\approx \mathit{elements}\times 1.44\times \log _2\left(\frac 1{\mathit{errorRate}}\right)
\end{equation*}
\begin{itemize}
\item Это формула является следствием от свойств фильтра Блума. Она действует для оптимального фильтра Блума. То есть,
для фильтра Блума с оптимальным числом функцией хеширования. О ней детально писали Pagh, Pagh, и Rao в 2005ом году (см.
списка литературы). Они в своей стати предлагают новую структуру, которая занимает меньше места (где фактор перед log —
единица), и которую мы бы тоже могли использовать. Это возможная оптимизация нашего подхода которую мы не рассмотрели.
\end{itemize}
\end{itemize}
\subsubsection[Время филировки и соединения]{Время филировки и соединения}
Зависимость времени последней фазы нашего алгоритма от процента ошибок, сделаны фильтром Блума — меньше простая. Однако,
последняя фаза выполняет множество разных задач вместе, благодаря автоматической генерации кода Apache Spark 2. Она
читает файл большой таблицы, и читает раздели маленькой таблицы из BlockManager, где они лежат с момента сформирования
фильтр Блума. Это фаза фильтрует записи с учётом условия нашего запроса и принадлежности к фильтру Блума. Затем, она их
сортирует, распределяет их в фазе shuffle и в конце концов, в последней стадии, она их соединяет, и пишет результат в
одном файле в формате parquet.

\paragraph{Формула времени последного задания программы, реализующей нашего подхода}
\begin{equation*}
\mathit{filterAndJoinTime}=L_1+L_2\times \mathit{errorRate}+\mathit{Poly}(\mathit{errorRate})\times \log
\left(\mathit{Poly}(\mathit{errorRate})\right)
\end{equation*}
где

\begin{itemize}
\item $L_1$ – постоянный элемент. Это время, которое платформа тратит на обработке записей нефильтрованной
большой таблицы, и на обработке элементов, которые есть в результате соединения (которые мы не должны отфильтровать).
Это включает, например, время чтения записей из файла на распределённой файловой системе. Мы назовём 
$N_{\mathit{filtrable}}$ число элементов большой таблицы, которые мы можем отфильтровать (которых нет в маленькой
таблице).
\item $L_2$ – линеарный фактор времени, затраченного на обработки элементов которые мы бы могли
отфильтровать, а которые фильтр Блума не смог найти. Число таких элементов —  $\text{э}\times N_{\mathit{filtrable}}$.
Этот линеарный фактор например соответствует времени обмена этих элементов между узлами через сеть, времени их записи
на диск, и времени пересортировки элементов с алгоритмом TimSort (см. Peters, Tim.
{\textquotedbl}Timsort.{\textquotedbl}, 2002).
\item  $\mathit{Poly}(X)=A\times X+B$. Этот полином смоделирует время обработки всех элементов, которые надо сортировать
в каждом разделе данных. Количество элементов в большой таблицы после фильтровки линейно зависит от процента ошибок
нашего фильтра Блума. 
$\mathit{count}(\mathit{LINEORDER}_{\mathit{filtered}})=\mathit{count}(\mathit{LINEORDER})-(1-\text{э})\cdot
N_{\mathit{filtrable}}$
\end{itemize}
\subsubsection{результаты модели}
 \includegraphics[width=\linewidth]{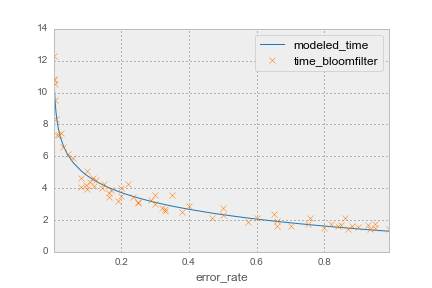} 

 \includegraphics[width=\linewidth]{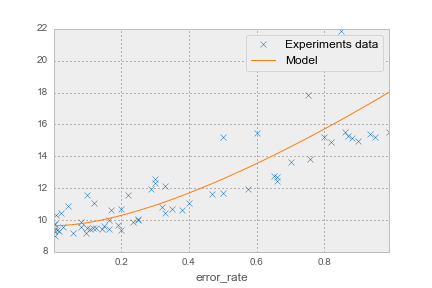} 

 \includegraphics[width=\linewidth]{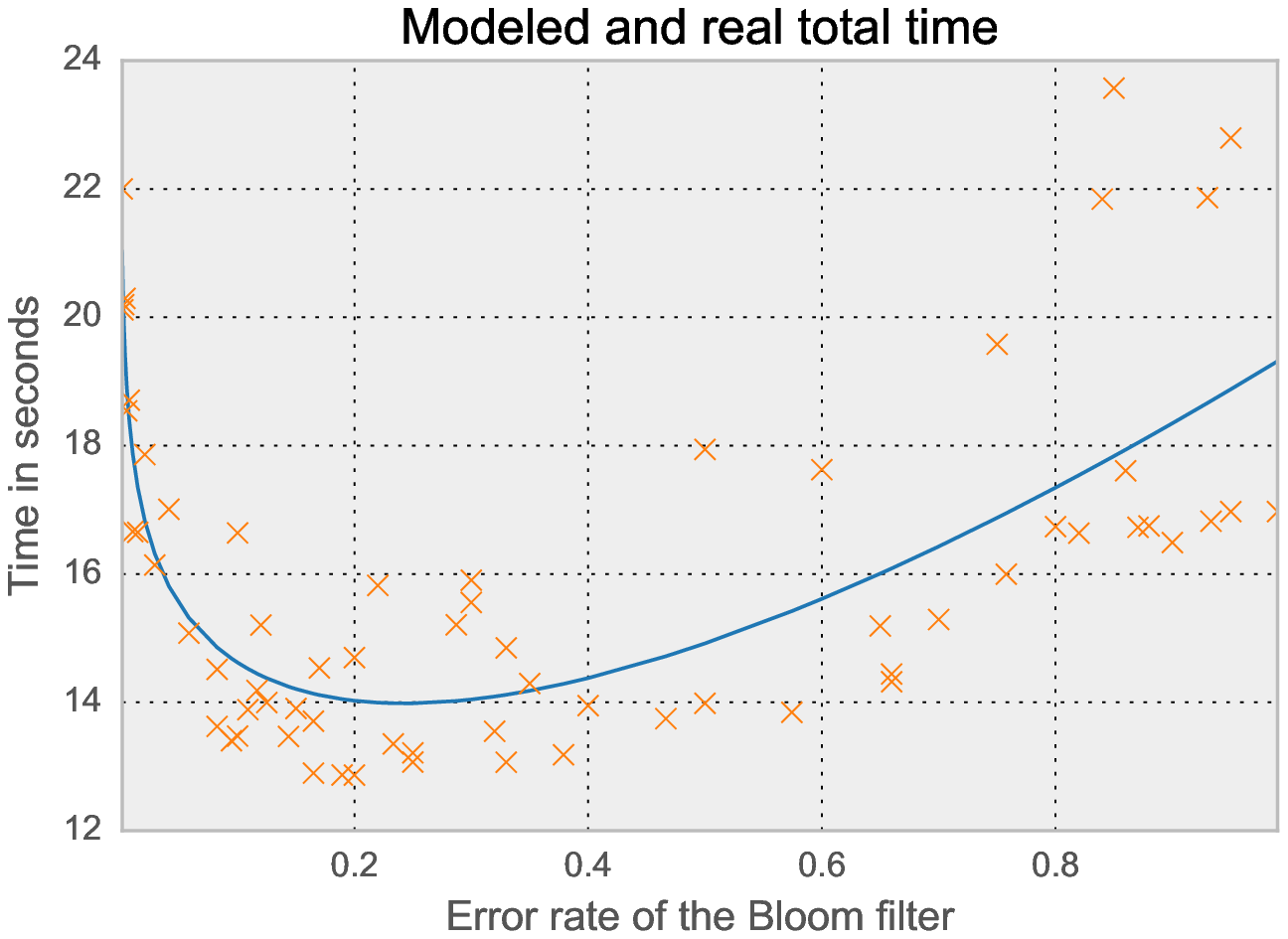} 

\subsection{Оптимизация времени соединения с помощью модели}

Модель тоталного времени - это просто сумма двух предыдуших моделей:

$$ model_{total}(\epsilon) = model_{bloom}(\epsilon) + model_{join}(\epsilon) $$
$$ model_{bloom}(\epsilon) = K_1 + K_2 \times log \left ( \frac 1 \epsilon \right ) $$
$$ model_{join}(\epsilon) = L_1 + L_2 \times \epsilon + (A \times \epsilon + B) \times \log(A \times \epsilon + B) $$

$$ \frac {d} {d\epsilon} (model_{bloom}(\epsilon)) = \frac{\partial}{\partial \epsilon}\left(K_{1} + K_{2} \log{\left (\frac{1}{\epsilon} \right )}\right) = - \frac{K_{2}}{\epsilon}$$
$$ \frac {d} {d\epsilon} (model_{join}(\epsilon)) = \frac{\partial}{\partial \epsilon}\left(L_{2} \epsilon + \left(A \epsilon + B\right) \log{\left (A \epsilon + B \right )}\right) = A \log{\left (A \epsilon + B \right )} + A + L_{2}$$

$$ \frac {d} {d\epsilon} (model_{total}(\epsilon)) =  A \log{\left (A \epsilon + B \right )} + A + L_{2} - \frac{K_{2}}{\epsilon}$$

То есть, мы можем найти оптимальный процент ошибок $\epsilon$, после
решения следующего уравнения:

$$ \begin{cases}
  \frac {d} {d\epsilon} (model_{total}(\epsilon)) =  0 \\
  0 < \epsilon \leq 1
  \end{cases}
$$
$$ \implies A \log{\left (A \epsilon + B \right )} + A + L_{2} - \frac{K_{2}}{\epsilon} = 0$$
$$ \implies \frac {K_2} {\epsilon} = A \log{\left (A \epsilon + B \right )} + A + L_2 $$

Символическое решение этого уравнения невозможное, но можно найти диапазон для  $\epsilon$ в зависимости от $A$, $B$, $L_2$ и $K_2$. Можно так же реализовать систему, которая оценила бы эти параметры до формирования филтра Блума, и решила бы уравнение численно. 
бы эти параметры до формирования филтра Блума, и решила бы уравнение численно. 

Эта система могла бы работать на главном узле, одновременно с первым заданием нашей программы, которое оценивает количество элементов в маленькой таблице. Она бы могла например использовать метод касательных Ньютона.

\section{Итоги}
В этой статье мы рассмотрели три основных аспекта. Мы рассмотрели новый алгоритм распределенного соединения данных с
помощью фильтра Блума, провели живые эксперименты на кластере и предложили простую математическую модель, позволяющую
выбрать оптимальное значение процента ошибок фильтра Блума. 

Следует заметить, что мы не затронули проблему детальной математической модели, которая бы приняла во входе больше
параметров кластера и которая бы позволила реализовать сразу в Spark оптимальную процедуру соединения. Подобная
процедура выбрала бы автоматически алгоритм соединения с фильтром Блума,  только если в том случае, если он выгоден и
выбрала бы оптимальные параметры фильтра.

\section{Список используемой литературы}
\begin{itemize}
\item
\href{http://www.lsi.upc.es/%7Ediaz/p422-bloom.pdf}{Space/Time Trade-offs in Hash Coding with Allowable Errors, Bloom, 1970} 
\item \href{http://oai.cwi.nl/oai/asset/21424/21424B.pdf}{\emph{\textup{TPCH ANalyzed, P. Boncz}}} 
\item \href{https://pdfs.semanticscholar.org/3fbf/943b7355d47b7c89a8631c1a2ed64d95dbfe.pdf}{Improving distributed join
efficiency with extended bloom filter operations, Loizos Michael at al.} 
\item \href{https://cacm.acm.org/magazines/2016/11/209116-apache-spark/fulltext}{Apache Spark: A Unified Engine for Big
Data Processing, Matei Zaharia et al., in Communications of the ACM, Vol. 59 No. 11, Pages 56-65}
\item \href{https://pdfs.semanticscholar.org/9bb8/48aa33cef74f760ca277978792bde8aac0c9.pdf}{Lee, Taewhi, Kisung Kim, and
Hyoung-Joo Kim. {\textquotedbl}Join processing using Bloom filter in MapReduce.{\textquotedbl} Proceedings of the 2012
ACM Research in Applied Computation Symposium. ACM, 2012.}
\item
\href{http://cirgs.org/wp-content/uploads/2015/09/Bloom-Filter-Based-Optimization-on-HBase-with-MapReduce.pdf}{Bhushan,
Mayank, Shashwati Banerjea, and Sumit Kumar Yadav. {\textquotedbl}Bloom filter based optimization on HBase with
MapReduce.{\textquotedbl} Data Mining and Intelligent Computing (ICDMIC), 2014 International Conference on. IEEE,
2014.}
\item \href{https://courses.cs.washington.edu/courses/cse544/11wi/projects/koutris.pdf}{Koutris, Paraschos.
{\textquotedbl}Bloom filters in distributed query execution.{\textquotedbl} (2011).}
\item \href{https://amplab.cs.berkeley.edu/wp-content/uploads/2015/03/SparkSQLSigmod2015.pdf}{Armbrust, Michael, et al.
{\textquotedbl}Spark sql: Relational data processing in spark.{\textquotedbl} Proceedings of the 2015 ACM SIGMOD
International Conference on Management of Data. ACM, 2015.}
\item \href{http://pages.cs.wisc.edu/~floratou/SQLOnHadoop.pdf}{Floratou, Avrilia, Umar Farooq Minhas, and Fatma Özcan.
{\textquotedbl}Sql-on-hadoop: Full circle back to shared-nothing database architectures.{\textquotedbl} Proceedings of
the }\href{http://pages.cs.wisc.edu/~floratou/SQLOnHadoop.pdf}{VLDB Endowment 7.12 (2014): 1295-1306.}
\item \href{http://www.academia.edu/download/36582865/grid05.pdf}{Bolze, Raphaël, et al. {\textquotedbl}Grid'5000: A
large scale and highly reconfigurable experimental grid testbed.{\textquotedbl} International Journal of High
Performance }\href{http://www.academia.edu/download/36582865/grid05.pdf}{Computing Applications 20.4 (2006): 481-494.}
\item Armbrust, Michael, et al. {\textquotedbl}Introduction to Spark 2.0 for Database Researchers.{\textquotedbl}
Proceedings of the 2016 International Conference on Management of Data. ACM, 2016.
\item \href{http://www.it-c.dk/people/pagh/papers/bloom.pdf}{Pagh, Anna, Rasmus Pagh, and S. Srinivasa Rao.
{\textquotedbl}An optimal Bloom filter replacement.{\textquotedbl} Proceedings of the sixteenth annual ACM-SIAM
symposium on Discrete algorithms. Society for Industrial and Applied Mathematics, 2005.}
\end{itemize}
\end{document}